\documentclass[twocolumn,apjl]{emulateapj}
\usepackage{natbib}
\usepackage{epsfig}
\usepackage{bm}
\usepackage{color}

\providecommand{\adsurl}[1]{\href{#1}{}}

\begin{document}

\newcommand{\clee}{C_{\ell}^{EE}}
\newcommand{\xef}{x_e^{\rm fid}}
\newcommand{\xet}{x_e^{\rm true}}
\newcommand{\dz}{\Delta z}
\newcommand{\zmax}{z_{\rm max}}
\newcommand{\zmin}{z_{\rm min}}
\newcommand{\zmid}{z_{\rm mid}}
\newcommand{\zre}{z_{\rm reion}}
\newcommand{\lcdm}{$\Lambda$CDM}
\newcommand{\wmap}{\emph{WMAP}}
\newcommand{\thetaa}{\theta_{A}}
\newcommand{\obhh}{\Omega_{b}h^2}
\newcommand{\ochh}{\Omega_{c}h^2}
\newcommand{\ascal}{A_{s}}
\newcommand{\nscal}{n_{s}}
\newcommand{\wh}[1]{\textcolor{blue}{(\bf #1)}}

\title{Reionization constraints from five-year WMAP data}
%\shorttitle{}

\author{Michael J. Mortonson$^{1,2}$ and Wayne Hu$^{1,3}$}
\affil{$^{1}$Kavli Institute for Cosmological Physics, 
Enrico Fermi Institute, University of Chicago, Chicago, IL 60637\\
$^{2}$Department of Physics,  University of Chicago, Chicago, IL 60637\\
$^{3}$Department of Astronomy and Astrophysics, University of Chicago, Chicago, IL 60637
}

\begin{abstract}
We study the constraints on reionization from five years of \wmap\ data, 
parametrizing the evolution of the average fraction of ionized hydrogen 
with principal components that provide a complete basis for describing the 
effects of reionization on large-scale $E$-mode polarization. 
Using Markov Chain Monte Carlo methods, we find that
the resulting model-independent estimate of the total optical depth 
is nearly twice as well determined as the estimate from 3-year \wmap\ data, 
in agreement with simpler analyses that assume instantaneous reionization. 
The mean value of the optical depth from principal components 
is slightly larger than the instantaneous value; we find 
$\tau=0.097\pm0.017$ using only large-scale polarization, and 
$\tau=0.101\pm0.019$ when temperature data is included.  Likewise,
scale invariant $n_s=1$ spectra are no longer strongly disfavored by WMAP alone.
Higher moments of the ionization history show less improvement in the 
5-year data than the optical depth. By plotting the distribution of 
polarization power for models from the MCMC analysis, we show that 
extracting most of the remaining information about the shape of the 
reionization history from the CMB requires better measurements of $E$-mode 
polarization on scales of $\ell\sim 10-20$.  Conversely, the quadrupole and octopole polarization
power is already predicted to better than cosmic variance given {\it any}
allowed ionization history
at $z<30$ so that more precise measurements will test the $\Lambda$CDM paradigm.
\end{abstract}


% =====================================================
\section{Introduction}
\label{sec:intro}

The amplitude of fluctuations in the $E$-mode component of 
cosmic microwave background (CMB)
polarization on large scales provides the current 
best constraint on the Thomson scattering optical depth to 
reionization, $\tau$. Assuming that the universe was reionized 
instantaneously, \cite{Dunetal08} estimate the total optical 
depth to be $\tau=0.087\pm0.017$ 
using five years of \wmap\ data.
Theoretical studies suggest that the process of reionization 
was too complex to be well described 
as a sudden transition~\citep[e.g.,][]{BarLoe01}.  
Previous studies have examined how the constraint on $\tau$ depends on 
the evolution of the globally-averaged ionized fraction during 
reionization, $x_e(z)$, for a variety of specific theoretical scenarios. 
If the assumed form of $x_e(z)$ is incorrect, 
the estimated value of $\tau$ can be biased; this bias can be lessened by 
considering a wider variety of reionization histories at the expense of 
increasing the uncertainty in $\tau$~\citep{Kapetal03,Holetal03,Coletal05}.

For the previous release of three years of \wmap\ data, using a more 
model-independent analysis of reionization based on principal components 
does not change the basic conclusions of studies that assume 
instantaneous reionization or other simple models; however, as the 
polarization power spectrum becomes better determined, it is 
increasingly important to adopt an approach with sufficient freedom to 
approximate a variety of possible reionization scenarios in order 
to minimize parameter biases \citep{MorHu08a,MorHu08b}. 
In this letter, we study how model-independent constraints on 
reionization have changed with the addition of two more years of 
data from \wmap\ and on what scales further measurements would have
the largest impact on ionization constraints. We review the principal component 
parametrization of the ionization history in \S~\ref{sec:pcs}. 
We then present the current constraints on the principal component 
amplitudes and the optical depth to reionization (\S~\ref{sec:tau})
and the range of polarization power spectra for general models that 
are currently allowed by the data (\S~\ref{sec:cl}). 
We discuss these results in \S~\ref{sec:disc}.

% ****************************************
\begin{figure*}[t]
\centerline{\includegraphics[width=3.5in]{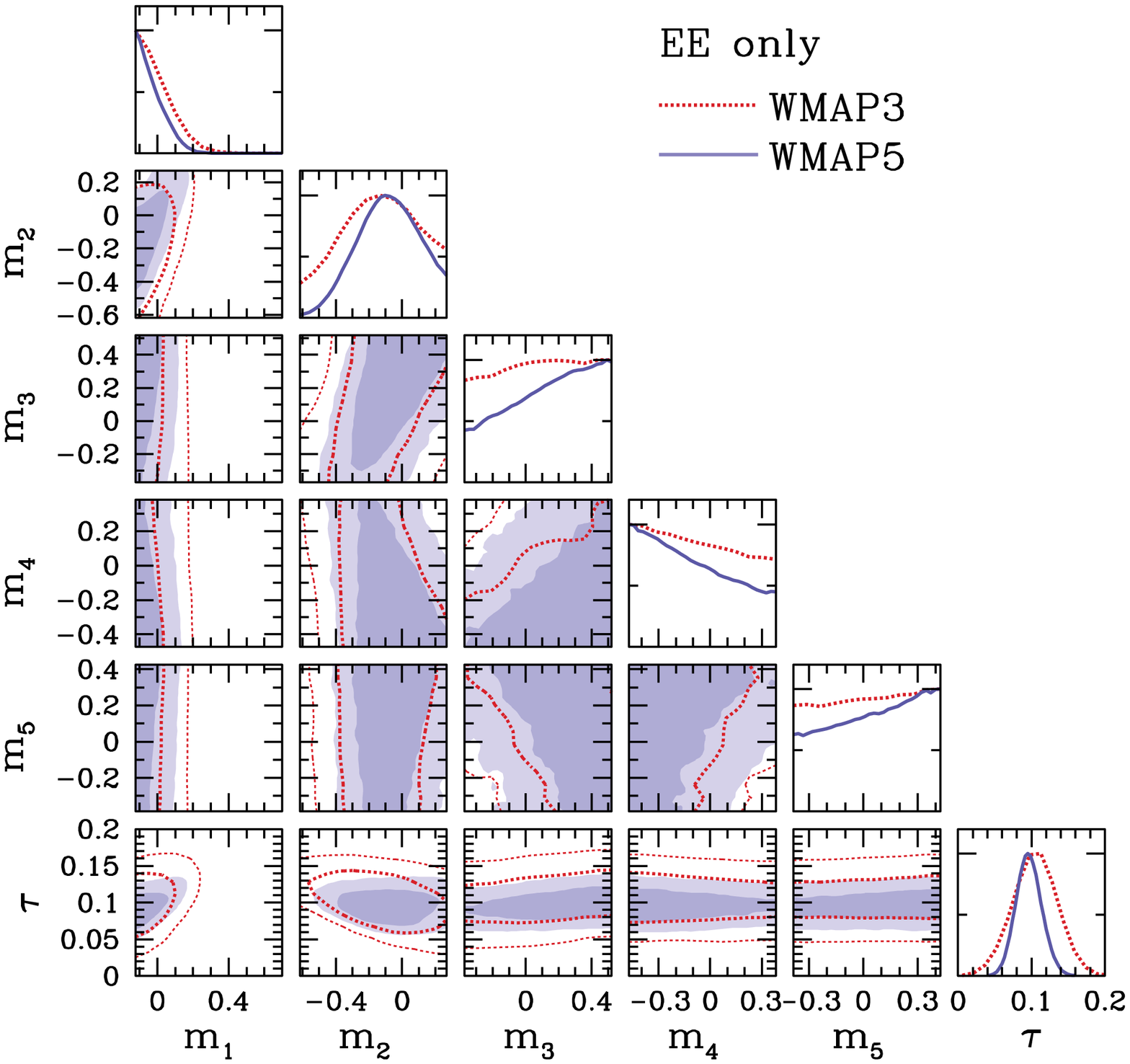}
\includegraphics[width=3.5in]{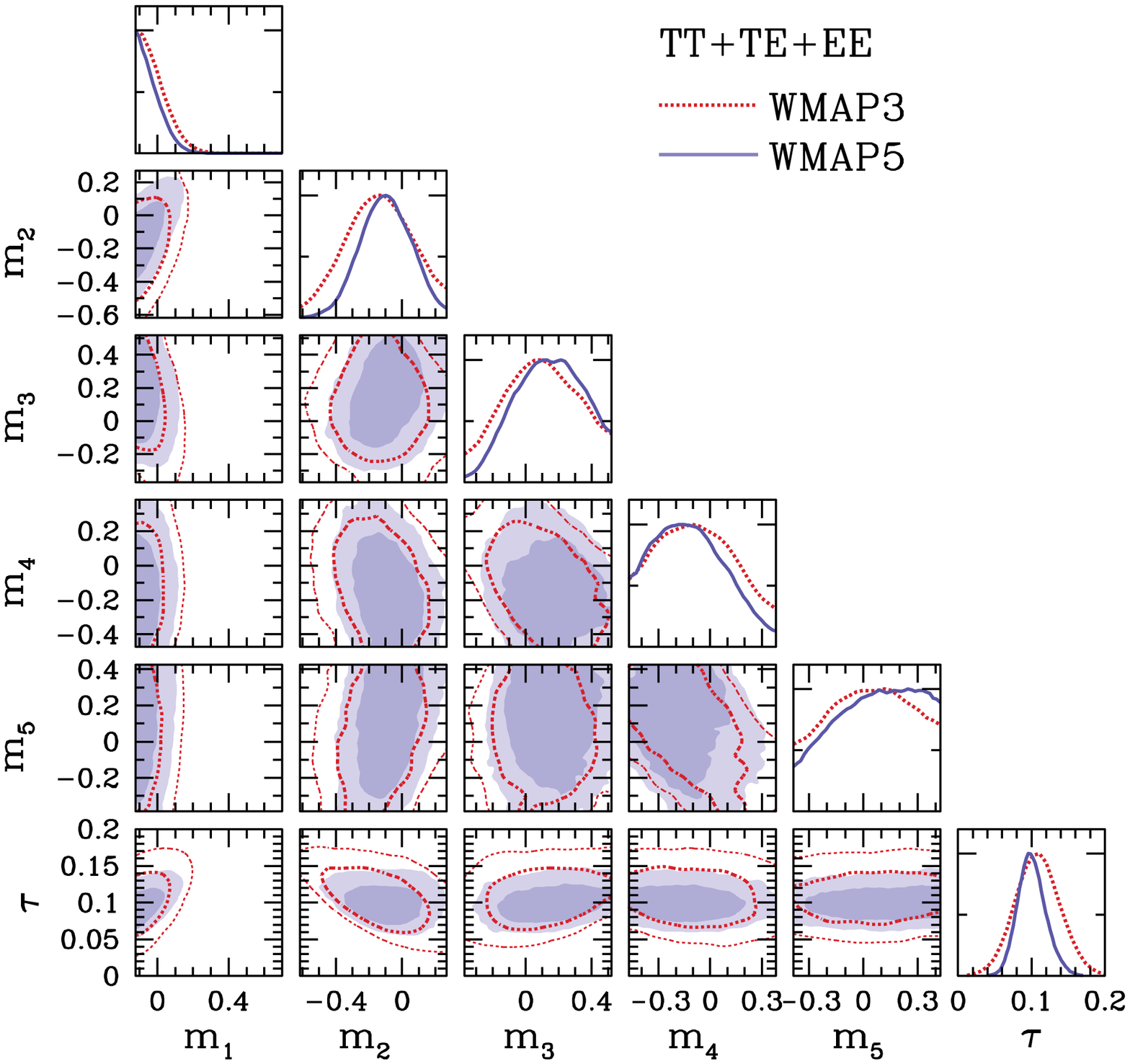}}
\caption{
Marginalized 2D 68\%  and 95\%  CL contours for the optical 
depth to reionization ($\tau$) and the amplitudes 
of the 5 lowest-variance principal components of $x_e(z)$ 
($m_{\mu},~\mu=1-5$). 
Panels along the diagonals show the 1D posterior probability 
distributions. Constraints are plotted for both 3-year 
(\emph{red dotted lines}) 
and 5-year (\emph{blue shading, solid lines}) \wmap\ data. 
In the left plot, only the 
low-$\ell$ reionization peak in the $E$-mode polarization power spectrum 
is used for parameter constraints, and all parameters besides the 5 
PC amplitudes and $\tau$ are held fixed. For the constraints in the right 
plot, we use both temperature and polarization data and allow five 
additional parameters to vary: $\obhh$, $\ochh$, $\thetaa$, 
$\ascal$, and $\nscal$.
The plot boundaries for the PC amplitudes correspond to physicality priors 
that exclude models that are unable to satisfy $0\leq x_e \leq 1$ for any 
combination of the higher-variance ($\mu \geq 6$) PCs.}
\vskip 0.25cm
\label{fig:pctau2d}
\end{figure*}
% ****************************************

% =====================================================
\section{Ionization Principal Components}
\label{sec:pcs}

We parametrize the reionization history 
as a free function of redshift by decomposing $x_e(z)$ into its
principal components (PCs) with respect to the $E$-mode polarization of the 
CMB~\citep{HuHol03,MorHu08a}:
\begin{equation}
x_e(z)=\xef(z)+\sum_{\mu}m_{\mu}S_{\mu}(z),
\label{eq:mmutoxe}
\end{equation}
where the principal components, $S_{\mu}(z)$, 
are the eigenfunctions of the Fisher matrix that describes 
the dependence of $\clee$ on $x_e(z)$, $m_{\mu}$ are the amplitudes of the 
principal components for a particular reionization history, and 
$\xef(z)$ is the fiducial model at which the Fisher matrix is computed.  The
components are rank ordered by their Fisher-estimated variances.
The lowest-variance eigenmode ($\mu=1$) is an 
average of the ionized fraction over the entire redshift range, weighted toward 
high $z$.  The $\mu=2$ mode measures the difference between 
the amount of ionization at high $z$ and at low $z$, and higher modes 
follow this pattern with weighted averages of $x_e(z)$ that oscillate with 
higher and higher frequency in redshift.
The main advantage of using principal components as a basis for $x_e(z)$ is 
that only a small number of the components are required to completely 
describe the effects of reionization on large-scale CMB polarization, 
so we obtain a very general parametrization of the reionization history at 
the expense of only a few additional parameters.

The principal components are defined over a limited range in redshift, 
$\zmin<z<\zmax$, with $x_e=0$ at $z>\zmax$ and $x_e=1$ at $z<\zmin$. 
We take $\zmin=6$, since the absence of Gunn-Peterson absorption in 
the spectra of quasars at $z\lesssim 6$ indicates that the universe is 
nearly fully ionized at lower redshifts \citep{FanCarKea06}. 
In the MCMC analysis presented here, we always use 
the five lowest-variance principal components of $x_e(z)$ with $\zmax=30$, constructed
around a constant fiducial model of
$\xef(z)=0.15$.  
The amplitudes of these components then serve to 
parametrize general reionization histories in the analysis of 
CMB polarization data. 
We refer the reader to~\cite{MorHu08a} for 
further discussion of these choices and the demonstration that five components 
suffice to describe the $E$-mode spectrum to better than cosmic variance precision.

We impose priors on the 
principal component amplitudes corresponding to physical values of the 
ionized fraction, $0 \leq x_e \leq 1$, according to the conservative
approach of \cite{MorHu08a}.  All excluded models are unphysical, 
but the models we retain are not necessarily strictly physical.
Finally, we neglect helium reionization, which is a small correction at the current 
level of precision but will be more important for 
future analyses \citep[e.g.,][]{ColPie08}.

% ****************************************
\begin{figure*}[t]
\centerline{\includegraphics[width=3.0in]{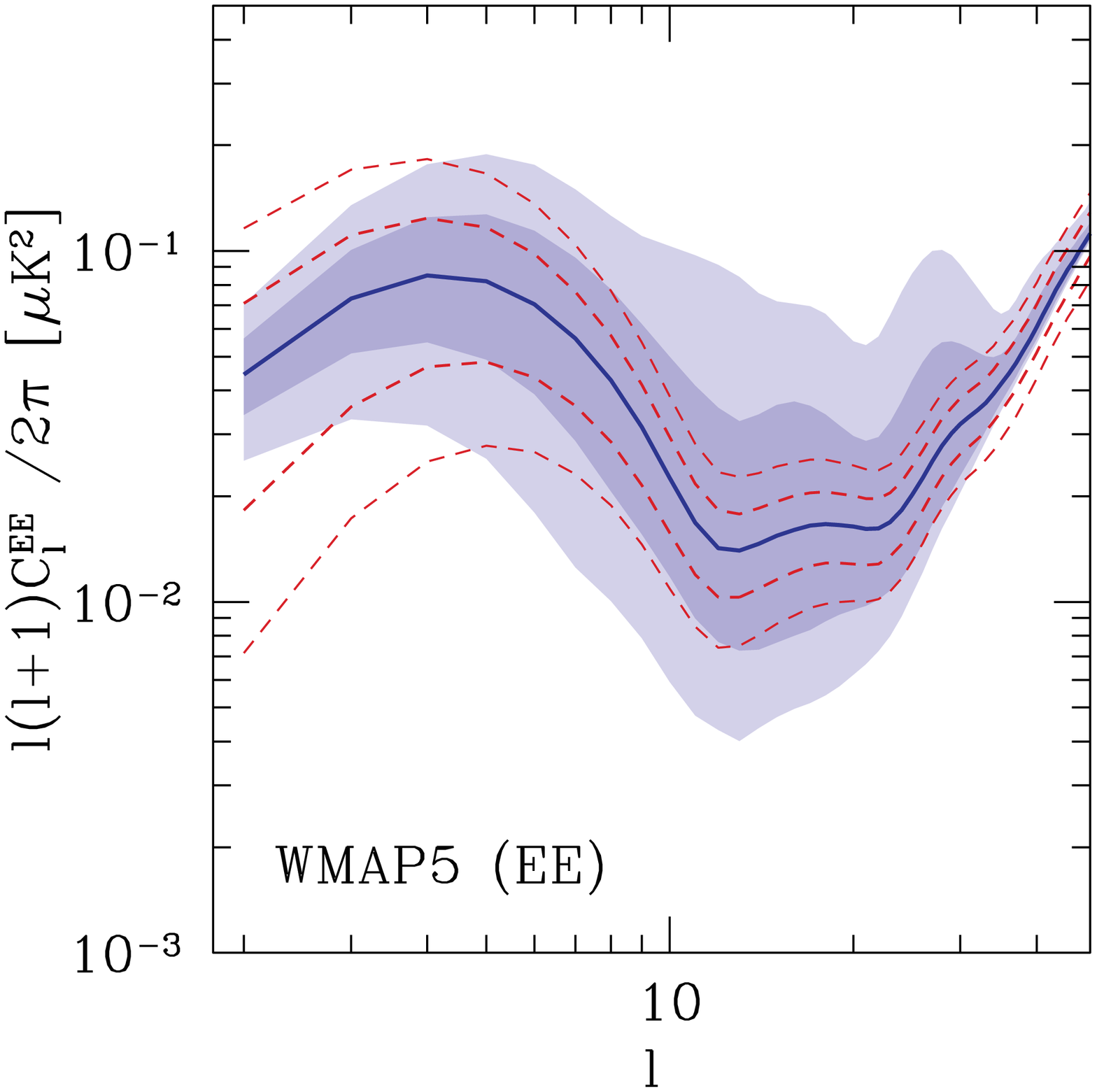}
\includegraphics[width=3.0in]{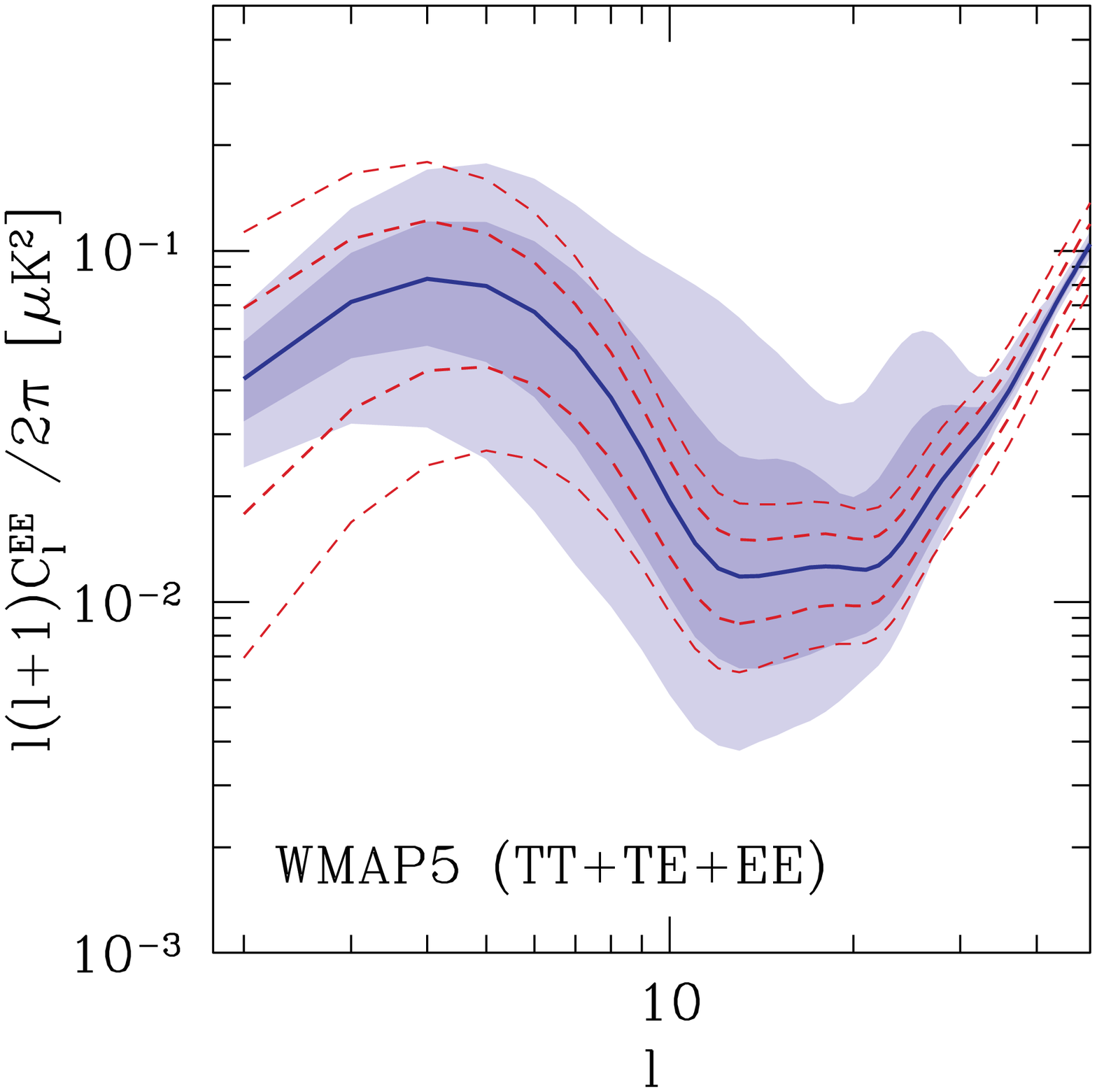}}
\caption{
Median (\emph{blue solid curve}) and 68\% and 95\% CL regions 
(\emph{blue shaded regions}) of polarization power spectra 
for any ionization history at $z<30$ allowed by the 5-year \wmap\ data,
assuming the standard $\Lambda$CDM paradigm. As in 
Fig.~\ref{fig:pctau2d}, only polarization data 
are used in the left panel and both 
temperature and polarization data are included in the right panel.
\emph{Red dashed curves}: cosmic variance (68\% and 95\% CL) around the median 
model.}
\vskip 0.25cm
\label{fig:cleedist}
\end{figure*}
% ****************************************

% =====================================================
\section{Optical Depth Constraints}
\label{sec:tau}

We examine the implications of the \wmap\ 5-year data for general 
models of reionization parametrized by principal components using a
Markov Chain Monte Carlo analysis that mirrors our previous study 
of the 3-year data in \cite{MorHu08a}. We consider constraints from 
either large-scale polarization alone, with parameters that do not directly 
affect reionization fixed to values 
that fit the temperature data (``EE''), or from the full set of 
temperature and polarization data, varying the parameters of the 
``vanilla'' \lcdm\ model (baryon density $\obhh$, cold dark matter density 
$\ochh$, acoustic scale $\thetaa$, scalar amplitude $\ascal$, and 
scalar spectral tilt $\nscal$) in addition to the reionization PC amplitudes 
(``TT+TE+EE''). In both cases, the total optical depth to reionization, 
$\tau$, is a derived parameter.

The MCMC constraints on principal component amplitudes and the 
derived optical depth for both of these cases are plotted 
in Fig.~\ref{fig:pctau2d}, along with the previous constraints 
from 3-year \wmap\ data \citep{MorHu08a}. While there are some 
improvements in all 5 of the individual components when considering $EE$ alone, 
these changes are not as large as the improvement in the optical 
depth constraint when all of the data are considered.  
 Adding both temperature data and extra parameters in going from EE only 
to TT+TE+EE has the net effect of slightly strengthening constraints on the higher ranked
PC amplitudes, although there is very little effect on $\tau$.
The additional constraining power for both 3-year and 5-year data 
comes mainly from the measured temperature power spectrum at 
$\ell\sim 10-100$, which excludes models with additional Doppler effect 
contributions due to narrow features in the ionization 
history \citep{MorHu08a}.

Modeling reionization as an instantaneous transition 
at some redshift $\zre$, \cite{Dunetal08} estimate the optical depth 
from the 5-year \wmap\ data to be $\tau=0.087\pm 0.017$, almost a factor 
of two more precise than the estimate from three years 
of data~\citep{Speetal07}.
For ionization histories parametrized by PCs, we find that the 
constraint on optical depth is $\tau=0.097\pm0.017$ for the EE case and 
$\tau=0.101\pm0.019$ for TT+TE+EE. As with the 3-year data, the error 
on $\tau$ is roughly 10\% larger with the inclusion of temperature data 
and a larger set of parameters, and the error in both cases is 
the same or only slightly larger than for the instantaneous reionization 
analysis. 

The central value of $\tau$ for more general ionization 
histories is higher than the instantaneous reionization value by 
$\sim 0.5-1~\sigma$; a similar shift toward larger optical depths was 
seen in PC analysis of the 3-year data \citep{MorHu08a}.  
The maximum likelihood model, however, has $\tau=0.088$ 
for EE data and $\tau=0.090$ for TT+TE+EE, much closer to the 
instantaneous reionization maximum likelihood optical depth of 
$\tau=0.089$ \citep{Dunetal08}.
The larger mean optical 
depth is at least partly due to having a large parameter volume
of models with finite ionization fraction
at high redshift that are still allowed by the data.   
With flat priors on the principal components, this volume effect
can boost the mean optical depth of models even though the mean likelihood
of low optical depth models remains the same.  Our assumption of 
full ionization at redshifts below $\zmin=6$ for all models also 
limits how small the optical depth can be. 
Relative to the best-fit instantaneous model 
with $\zre=11.0\pm 1.4$~\citep{Dunetal08}, there are simply more ways 
to increase $\tau$ than there are to decrease $\tau$ by changing the 
ionization history, given these priors and the current data.

For the TT+TE+EE analysis, the larger mean optical depth 
is accompanied by shifts in correlated parameters, particularly 
the spectral tilt: $n_s=0.990\pm0.024$ with $x_e(z)$ parametrized by PCs, 
and $n_s=0.960\pm0.015$ for instantaneous reionization \citep{Kometal08}. 
As with the optical depth, however, some of this shift is a parameter 
volume effect. The maximum likelihood model for the principal component 
analysis has $n_s=0.976$. The best fit scale invariant model  
(fixing $n_s=1$) is a poorer fit to the data by 
$\Delta \chi_{\rm eff}^2 \equiv -2 \ln (\mathcal{L}/\mathcal{L}_{\rm max}) \sim 1$, where $\mathcal{L}_{\rm max}$ is the maximum likelihood.
(The instantaneous reionization maximum likelihood model is also at 
$\Delta \chi_{\rm eff}^2 \approx 1$ relative to the best fit with principal 
components.)
As measurements of CMB polarization improve with future data, particularly
with detections in the $10 < \ell < 20$ range (see \S \ref{sec:cl}), 
the constraints on parameters such as optical depth and tilt should 
become less sensitive to our assumptions about the priors.

Unlike the total optical depth, constraints on the optical depth 
over more limited redshift ranges have only improved slightly. 
With three years of \wmap\ data, the 95\% upper limit on 
the optical depth from $z>20$ (allowing for a significant ionized 
fraction up to $z\sim 40$) was $\tau(z>20)<0.08$ \citep{MorHu08a}. The 
limit from 5-year data is $\tau(z>20)<0.07$. 
If we instead choose the dividing redshift to be the best-fit value of 
the redshift of instantaneous reionization, $\zre=11$, we find 
a similar constraint for the contribution to the optical depth 
from high redshift: $\tau(z>11)<0.07$. Compared to the 3-year data, 
there is also a more significant (but still weak) preference 
for nonzero optical depth from $6<z<11$.

% =====================================================
\section{Power spectra of allowed models}
\label{sec:cl}

To better understand at which scales the reionization peak of 
$E$-mode polarization is best constrained by the current data, 
we plot the 68\% and 95\% CL limits on $\clee$ from the Monte Carlo chains in 
Fig.~\ref{fig:cleedist}.  These limits reflect the range of
ensemble-averaged power allowed by the 5-year data and
the 
PC-parametrized reionization histories.
Since this parametrization is complete in the power spectrum, 
the range in Fig.~\ref{fig:cleedist} reflects
the allowed model power spectra for {\it any} ionization history at
$z<z_{\rm max}=30$.

At $\ell\lesssim 5$, the variation in allowed models is smaller 
than the uncertainty due to cosmic variance.   In other words,
the data at $\ell \sim 5$ in combination with any ionization history and
the power law initial power spectrum
make a prediction for the ensemble-averaged power at lower $\ell$ that is
sharper than can be measured.    Conversely, measurements that
violate this prediction at a statistically significant level require modifications to
 the
$\Lambda$CDM paradigm itself, much like low measurements of the temperature
quadrupole.  It is interesting that the maximum likelihood $E$-mode polarization
quadrupole reported by \cite{Noletal08}, 
$6 C_2^{EE}/2\pi \approx 0.15~\mu{\rm K}^2$, is in excess of the 95\% cosmic
variance region shown in Fig.~\ref{fig:cleedist}.

The uncertainty in the model space is largest at intermediate scales 
of $\ell \sim 10-20$, where the large-scale polarization power is expected to 
be smallest. There is substantial room for improved measurements of 
the spectrum on these scales before reaching the cosmic variance limit.
Tighter constraints on the $E$-mode power at $10<\ell <20$ would 
better determine the amplitude of principal components beyond the first; 
such measurements are necessary to be able to discriminate among 
different reionization histories with the same total optical depth.
Physically, these measurements would better constrain the ionization 
history at high redshifts ($z\gtrsim 15$).

On small scales ($\ell > 30-40$) the limits on power spectra of 
reionization models from the chains again become tighter than 
cosmic variance since the theoretical amplitude of the 
recombination peak is well determined due to constraints on 
parameters from the temperature data.

Comparison of the two panels in Fig.~\ref{fig:cleedist} shows that 
the main effect of including temperature data in constraints on 
principal components is to eliminate models with large power at 
$10<\ell<30$.  As mentioned in the previous section, these models are 
excluded by the data due to their increased temperature fluctuations at 
$\ell \sim 10-100$.
Even with TT+TE+EE data, the range of power in 
models allowed by current data is a few times larger than 
cosmic variance.

% =====================================================
\section{Discussion}
\label{sec:disc}

The 5-year \wmap\ polarization data significantly improve the 
estimate of the total optical depth, reducing the error from 
$\sigma_{\tau}\approx 0.03$ to $\sigma_{\tau}\approx 0.017$. 
This improvement is seen in both a model-independent analysis 
using principal components of the ionization history and in 
an analysis that assumes instantaneous reionization, although there 
is a small shift in the central value with the model-independent 
method preferring a slightly higher mean around $\tau=0.1$.

As with the 3-year data, the $E$-mode reionization peak is currently 
best measured on the largest scales, $\ell\sim 5$. Determining 
details of the ionization history beyond the optical depth requires 
information about the full shape of the reionization peak, which 
can be obtained by supplementing the current observations with 
better measurements of the $E$-mode power on scales of $5<\ell<30$.
In particular, improved knowledge of the power on scales between 
the main reionization peak and the recombination peak at $\ell\sim 10-20$ 
would be the most useful for distinguishing models of 
reionization with different ionization histories but the same optical depth.

Conversely, the data along with {\it any} allowed ionization history at $z<30$ 
in the standard $\Lambda$CDM context already predict the 
ensemble-averaged polarization quadrupole
and octopole powers to better than cosmic variance.   More precise measurements
in this regime can test the standard model itself.

\acknowledgements {\it Acknowledgments}:  
%restore if we come up with list: We thank  ... for useful discussions.  
This work was supported by the KICP through the grant NSF PHY-0114422 and
the David and
Lucile Packard Foundation.
WH was additionally supported by  the DOE through
contract DE-FG02-90ER-40560.

\bibliographystyle{apj}
%\bibliography{wmap5pc}

\end{document}